\documentclass[aps,prb,preprint]{revtex4}

\usepackage{graphicx}

\usepackage{amsmath}

\begin{document}

\title{On the macroscopic verifications of Klein's theorem and the proof of $E_0=mc^2$}

\author{T. C. Choy}
\email{tuckvk3cca@gmail.com}
\affiliation{London Centre for Nanotechnology and Department of Physics and Astronomy, University College London, Gower Street, London WC1E 6BT, UK}

\begin{abstract}
Alternative verifications of Klein's theorem and the proof of $E_0=mc^2$, for a relativistic macroscopic body are presented, using models with boundary conditions of varying complexity, together with some refinements for the case containing electromagnetic radiation for the simplest model. The robustness of these models to the final result of $E_0=mc^2$, attests to the minor role played by the Poincar\'e type stresses introduced in some of these models for mechanical stability. Finally we caution the reader that while internal consistency of the $E_0=mc^2$ relation for a macroscopic body in special relativity is proved, it does not in any way furnish a proof of the relation for a single point particle, for this would imply that one is able to prove the postulates of special relativity from the premises of the theory itself.

\end{abstract}

\maketitle

\section{Introduction}
\label{Introduction}

In a recent article, Hans Ohanian\cite{Ohanian1}, presented a general proof of Einstein's $E_0=mc^2$ as a corollary of a mathematical theorem for conserved energy-momentum tensors, by  Felix Klein in 1918 \cite{Klein1918}. It is true that Klein's theorem, also known as Abraham's theorem \cite{Holstein}, is often only mentioned in passing and their contributions had been largely unacknowledged or forgotten in history in most standard textbooks, see for example \cite{Landau1, PanofskyandPhilips1} and others \cite{Barut, Jackson}. In this short note we shall examine alternative verifications of the Klein-Abraham theorem and therefore $E_0=mc^2$,  from the perspective of (non-stationary) relativistic macroscopic models. An example of this is the famous parallel plate capacitor, cf Comay\cite{Comay}, which will demonstrate the constraints on the energy momentum tensor (over the {\it localized} region where it is diagonalizable) due to the conservation law. This material should be included in standard texts as it is fairly simple.  In addition the case involving radiation requires some refinements which shall also be presented here for the simplest model.  This note supplements that of earlier authors for which the energy momentum tensor had already been discussed in some detail in this journal over the last two decades or more, including several contributions from the ex-editor Romer\cite{Romer1} and others, see for example \cite{Holstein, Rohrlich1,MacDonald,Comay,Comay1,Hnizdo1,Hnizdo2,Hnizdo3} and references cited therein.
In addition, the robustness of the final result $E_0=mc^2$, is an attestation of the covariant prescription for the external field, originally due to Fermi and later championed by Rohrlich \cite{Rohrlich0,Moylan} and thus for most purposes, it pays to choose the simplest of boundary conditions as shown here. Our motivations here are two fold, (a) to provide a picture of how the Klein-Abraham theorem operates macroscopically which is suitable for undergraduate teaching and (b) to provide an approach for radiation missing in Ohanian's paper \cite{Ohanian1} since in this case the energy momentum four vector is a null vector, and the latter's arguments in the paragraph containing his equation(8) are no longer valid. We conclude with some open questions to stimulate further research.

\section{Rest frame energy-momentum tensor of a unit volume}
\label{section1}
The starting point is to note that there exists an arbitrary unit volume $V_{AB}$ of a macroscopic body which is a rest frame $K_0$ relative to other parts of the body which can have macroscopic velocities comparable to $c$. However before building up the body from such elemental constituents, we shall consider an elementary model such as the parallel plate capacitor in its rest frame (which is globally diagonalizable). A detail analysis with a homogeneous dielectric has already been given by Comay in this journal more than a decade ago \cite{Comay}. For a material system (such as the dielectric media in the capacitor) that is constrained by the speed of light, it can have only one form \cite{Note1,Landau2}:

\begin{equation}
T^{(0)\alpha\beta}=
\begin{pmatrix}
\epsilon_{0}&0&0&0\cr
0&\epsilon_{1}&0&0\cr
0&0&\epsilon_{2}&0\cr
0&0&0&\epsilon_{3} \\
\end{pmatrix},
\label{equation1}
\end{equation}
where $\epsilon_{0}$ is the energy density and $\epsilon_{i}, (i=1\ {\rm to}\ 3) $ are the components of the pressure density at the four dimensional spacetime point $x^\alpha$ and are the invariants of the tensor. Such a region need not be global as long as it is essentially macroscopic (or mesoscopic to use a modern terminology), so that our unit volume is an elementary representative of the entire body, whose other parts are determined via a proper Lorentz transformation, see later. In spite of their simplicity, our models are not static as in Laue's theory, its microscopic constituents can in fact be a relativistic gas in motion or a standing electromagnetic wave (whence $\epsilon_{\alpha}$ becomes time dependent). The mutual consistency between the microscopic and macroscopic descriptions has been shown by Comay \cite{Comay} and we shall not discuss this any further here.

Without loss of generality, for a frame $K$ in which $K_0$ is moving with velocity $v$ relative to it along the positive $x^0$ axis, the tensor Eq(\ref{equation1}) can be obtained by a proper Lorentz transformation on a second rank symmetric tensor as, see for example \cite{Comay,Landau3}:
\begin{equation}
T^{\alpha\beta}= \gamma^2
\begin{pmatrix}
(\epsilon_{0}+\beta^2\epsilon_{1}) &\beta(\epsilon_{0}+\epsilon_{1})&0&0\cr
\beta(\epsilon_{0}+\epsilon_{1})&(\epsilon_{1}+\beta^2\epsilon_{0})&0&0\cr
0&0&\epsilon_{2}&0\cr
0&0&0&\epsilon_{3} \\
\end{pmatrix},
\label{equation2}
\end{equation}
where $\beta=v/c$ and $\gamma^{-1}=\sqrt{(1-\beta^2)}$ as usual.  Note that without the constraints imposed by the conservation laws,  the integrated quantities $P^{\alpha}=(P^0,{\bf P})$ do not transform as true four vectors in general, as integrating over the unit volume gives:
\begin{eqnarray}
P^0 = {1\over c}\int d{\bf r} T^{00}  &=& {\gamma^2\over c} \int d{\bf r} [\epsilon_{0}+\beta^2\epsilon_{1}] \ {\rm while}\cr
P^1= {1\over c} \int d{\bf r} T^{01}  &=& {\gamma^2\beta \over c}\int d{\bf r} [\epsilon_{0}+\epsilon_{1}].
\label{equation3}
\end{eqnarray}
Following Comay \cite{Comay} we could call such quantities false four vectors.  However, if the conservation law holds, they impose constraints given by:
\begin{equation}
{\partial T^{(0)\alpha\beta}\over \partial x^{\beta} }= 0,
\label{equation4}
\end{equation}
which dictate that:
\begin{eqnarray}
{1\over c}{\partial \epsilon_0 \over \partial t} &=& 0,\  \ {\rm and} \cr
{\partial \epsilon_i \over \partial x^i} &=& 0\ (i=1,2,3),
\label{equation5}
\end{eqnarray}
where no summation over $i$ is implied in the last equation. Notice the decoupling of the space and time components here which is a great advantage obtained by diagonalization. Firstly, a volume integration of the first of equation(\ref{equation5}) over the rest frame unit volume element (i.e. $d{\bf r_0}$) shows that the total energy $E_0 = \int d{\bf r_0}  \epsilon_0$ is a constant of motion, i.e. independent of rest frame time $t_0$. Next a similar integration of the last of equation Eq(\ref{equation5}) {\it requires} that the net force components cancel across the two boundaries A and B:
\begin{equation}
\int d{\bf r_0} {\partial \epsilon_i \over \partial x^i} =
\int d{S_A}\epsilon_i(x_A)-\int d{S_B}\epsilon_i(x_B) = 0,
\label{equation5a}
\end{equation}
which is a condition for mechanical equilibrium.  At this point the problem reduces to appropriate choices for the boundary conditions.  We shall look at some simple cases in the next subsections.
\subsection{Uncharged Capacitor}
The simplest case in which equation(\ref{equation5a}) can be satisfied is the vanishing of all stress components $\epsilon_i(x_A)=\epsilon_i(x_B)=\epsilon_i(x_C)=0$, such as the dielectric between the plates of an uncharged capacitor in the model of Comay\cite{Comay}. Since the derivatives must also vanish throughout the body, then  $\epsilon_{i}=0$ for $({\rm i}=1,2,3)$ is required by conservation of the energy momentum tensor.  There is thus no pressure in this simplest of models, unless we have radiation (see later).  We can now, following Hnizdo \cite{Hnizdo2}, integrate  equation (\ref{equation3}) and change the variable of integration from ${\bf r}(t)$  at time $t$ to ${\bf r}_0(t)$  at the same time $t$. The time $t$ corresponds to a different rest frame time $t_0$, but that does not matter as the energy is conserved in the rest frame time $t_0$. Thus equation (\ref{equation3}) becomes: \begin{eqnarray}
P^0 = {\gamma^2\over c} \int d{\bf r}\epsilon_0&=& {\gamma\over c} \int d{\bf r}_0\epsilon_0= {\gamma \over c}E_0 \cr
P^1 ={\gamma^2 \beta\over c}\int d{\bf r} \epsilon_0&=&{\gamma \beta\over c}\int d{\bf r}_0\epsilon_0= {\gamma \beta \over c} E_0,
\label{equation6}
\end{eqnarray}
which verifies Klein's and Abraham's theorem, Einstein's signature relation $E_0=m c^2$ and Ohanian's equation(8) \cite{Ohanian1}.

\subsection{Charged Capacitor}
\label{section B}
For the case of a charged capacitor, the stress components $\epsilon_i$ do not vanish. In this case the pressure has to be balanced by an opposite pressure from other means, either via forces applied externally or via an electric field $E$ between the capacitor plates \cite{NoteA} taken as the $x$ axis. A proper analysis will have to include the electromagnetic energy-momentum tensor due to these static fields. Equation(\ref{equation1}) now becomes \cite{Comay} :
\begin{equation}
T^{(0)\alpha\beta}=
\begin{pmatrix}
\epsilon_{0}+W_E&0&0&0\cr
0&\epsilon_{1}-W_E&0&0\cr
0&0&\epsilon_{2}+W_E&0\cr
0&0&0&\epsilon_{3}+W_E \\
\end{pmatrix},
\label{equation1a}
\end{equation}
where $W_E={E^2\over 8\pi}$, and the field $E$ is taken along the $x$ axis. Equation(\ref{equation3}) now becomes:
\begin{eqnarray}
P^0 = {1\over c}\int d{\bf r} T^{00}  &=& {\gamma^2\over c} \int d{\bf r} [(\epsilon_{0}+W_E)+\beta^2(\epsilon_{1}-W_E)] \ {\rm while}\cr
P^1= {1\over c} \int d{\bf r} T^{01}  &=& {\gamma^2\beta \over c}\int d{\bf r} [(\epsilon_{0}+W_E)+(\epsilon_{1}-W_E)].
\label{equation3a}
\end{eqnarray}
Now the second integrals in the above equations can be transformed using the following trick:
\begin{eqnarray}
\int d{\bf r} (\epsilon_{1}-W_E)  &=& \int d{\bf r} {\partial [x (\epsilon_{1}-W_E)]\over \partial x} \ -\int d{\bf r} x f_x\ \cr
  & = & \oint dS_A [x (\epsilon_{1}-W_E)]|_{x=x_A}.
\label{equation3b}
\end{eqnarray}
where $f_x={\partial \epsilon_1 \over \partial x}$ is the stress density component normal to that surface.
The integral over $f_x$ vanishes (as can be seen by multiplying the last of equation(\ref{equation5}) by $x$ and integrating over the volume). Now the surface integrals can be arranged to vanish by the surface charges $(\epsilon_{1}-W_E)|_{x=x_A}=0$ at the boundary \cite{Comay}, but once again the conservation law (see equation (\ref{equation5})) requires the derivative to vanish, so $(\epsilon_{1}=W_E)$ in the bulk as well.  The remainder of the arguments of the last subsection follow through verifying once again Einstein's signature relation $E_0 + E_{W}=m c^2$, where we must now include the energy of electrostatic field $E_{W}=\int d{\bf r} W_E$.

The electrostatic field $E$ actually plays no significant role other than to provide mechanical equilibrium in what we are concerned with next, in particular their cross-coupling with the radiation fields vanish on average, see Comay \cite{Comay2}. We could therefore ignore them, bearing in mind that they do exist in the background if the capacitor is charged.  For simplicity we shall instead continue to consider an uncharged capacitor of unit volume but include radiation, in this case a pressure will arise from radiation that has to be balanced mechanically by applying an external force on the plates, see below and later.

\section{Unit volume uncharged capacitor with radiation}
\label{CapacitorwithRadiation}
Following the above approach the inclusion of radiation \cite{Note2} can be treated similarly, but here the energy momentum tensor consists of two parts, $T^{\alpha\beta}_{total}=T^{\alpha\beta}_{body} + T^{\alpha\beta}_{rad}$ and it is the total that is conserved, not the individual parts \cite{Landau4, Holstein}. Moreover, due to the fact that $c$ is the maximum velocity of propagation of energy flux, and the pseudo-Euclidean nature of Minkowski spacetime, the tensor $T^{\alpha\beta}_{rad}$ cannot be diagonalized \cite{Landau2} i.e. there can be no rest frame for light.  Consequently the arguments of the previous section are in need of some refinements. However, it is still convenient to start in the rest frame of the body for which equation (\ref{equation1}) is retained for $T^{\alpha\beta}_{body}$.  For the electromagnetic part, we shall consider without loss of generality, propagation of plane waves in the positive $x^0$ direction only, for which:
\begin{equation}
T^{\alpha\beta}_{rad}=
\begin{pmatrix}
W&W&0&0\cr
W&W&0&0\cr
0&0&0&0\cr
0&0&0&0 \\
\end{pmatrix},
\label{equation7}
\end{equation}
where $W$ is the electromagnetic energy density: $W={1 \over 8\pi}(E_y^2 +H_z^2)={1 \over 4\pi}E_y^2={1 \over 4\pi}H_z^2$, in terms of the electric $E$ and magnetic $H$ fields, see for example case (c) of reference \cite{Landau2}.  Now the total tensor takes the form:
\begin{equation}
T^{\alpha\beta}_{tot}=
\begin{pmatrix}
\epsilon_0+W&W&0&0\cr
W&\epsilon_1+W&0&0\cr
0&0&\epsilon_2&0\cr
0&0&0&\epsilon_3 \\
\end{pmatrix},
\label{equation8}
\end{equation}
which can be diagonalized.  The diagonalized tensor which corresponds to the centre of inertia frame now takes the form \cite{Note3}:
\begin{equation}
T^{(0)\alpha\beta}_{tot}=
\begin{pmatrix}
\lambda_0&0&0&0\cr
0&\lambda_1&0&0\cr
0&0&\epsilon_2&0\cr
0&0&0&\epsilon_3 \\
\end{pmatrix},
\label{equation9}
\end{equation}
where the roots are given by:
\begin{equation}
\lambda_0={1\over2} [(\epsilon_0-\epsilon_1) + \sqrt{(\epsilon_0-\epsilon_1)^2+4(\epsilon_0\epsilon_1+W(\epsilon_0+\epsilon_1))}],\
\label{equation10}
\end{equation}
and
\begin{equation}
\lambda_1={1\over2} [-(\epsilon_0-\epsilon_1) + \sqrt{(\epsilon_0-\epsilon_1)^2+4(\epsilon_0\epsilon_1+W(\epsilon_0+\epsilon_1))}].
\label{equation11}
\end{equation}

The ability to diagonalized the energy-momentum tensor provided by the material body enables the exercise of the arguments of the previous section.  Here, $\lambda_1$ must vanish, (in addition to $\epsilon_2$ and $\epsilon_3$) and thus:

\begin{equation}
\epsilon_1=-{\epsilon_0 W\over \epsilon_0+W},
\label{equation12}
\end{equation}

\noindent is the extra pressure density that must be applied externally to balance that due to radiation.  Using equation(\ref{equation10}), the mass increase $\Delta m = m_W-m$ due to radiation can be obtained, by employing results of the previous section as:
\begin{equation}
\Delta m={1\over c^2} \int d{\bf r}_0 \epsilon_0 [{W\over \epsilon_0+W}],
\label{equation13}
\end{equation}
where the integration is carried out in the centre of inertia frame. To leading order in $W^2/\epsilon_0$, the mass increase $\Delta m=U_{rad}/c^2$, in terms of the integrated radiation energy $U_{rad}$.  Equation(\ref{equation13}) shows that $\Delta m/m$ saturates to unity as $W \rightarrow \infty$.  Note that the mass-energy formula for radiation holds here even though $T^{\alpha\beta}_{rad}$ is not conserved.  As pointed out by Rohrlich \cite{Rohrlich1}, the latter is a sufficient, and not a necessary condition. For an account of the erroneous $4/3$ factor discrepancy \cite{PanofskyandPhilips2}, see references \cite{Kwal,Rohrlich2,Moylan}.  In the next section we shall conclude by taking up the problem of the full macroscopic body which is not globally diagonalizable but we will not consider radiation any further apart for some comments at the end.

\section{Full macroscopic body}
The above apparatus allows us to consider the full macroscopic body in the following way.  As long as the unit volume element of section \ref{section1} is {\it locally} diagonalizable in the form of equation(\ref{equation1}),  the full energy-momentum tensor is now given by \cite{Landau2}:
\begin{equation}
T^{\alpha\beta}= (p+\epsilon_0) u^{\alpha}u^{\beta}-pg^{\alpha\beta},
\label{equation14}
\end{equation}
where for convenience we have assumed the validity of Pascal's law i.e. $\epsilon_i=p, i = 1\ {\rm to}\ 3$, since it holds in fluids and the maximum possible deviations from it are in general relativistically insignificant\cite{Landau5}. In this description, we can see that local elements are in general non-interacting and differ from each other only via an arbitrary Lorentz boost, subject only to continuity and other boundary conditions imposed by energy-momentum conservation as shown in section \ref{section1}.  \subsection{Stressless body}
\label{stressless body}
We shall follow the last section and consider the simplest model with $p=0$ first, i.e. an unstressed body and therefore its constituents cannot be subjected to rotational motion.
Now, if our unit volume elements are to be locally diagonalizable, then the velocity field must be considered constant over that local element which we shall call a unit cell.  Hence a volume integral of the tensor equation(\ref{equation14}) with $p=0$ can be partitioned:
\begin{equation}
\int d{\bf r} T^{\alpha\beta} = \sum_j \int{d{\bf r}_{(j)}} \epsilon_0 u_{(j)}^{\alpha}u_{(j)}^{\beta}\ \ ,
\label{equation15}
\end{equation}
 where the integrals in the sum are over each unit cell $j$.  The key to what follows is that the velocities can all be taken out of the integration in each of these elements, further to our comments above. Let us now perform these integrations in the centre of inertial frame (to be defined below) and see how it works. Now the integrals reduce to:
\begin{equation}
\int d{\bf r} T^{\alpha\beta}= \sum_{j} u_{(j)}^{\alpha}u_{(j)}^{\beta} \int d{{\bf r}_{(j)}}  \epsilon_0=\sum_{j} \gamma^{-1}_{(j)} u_{(j)}^{\alpha}u_{(j)}^{\beta} E_0=V E_0\langle\gamma^{-1}_{(j)}u_{(j)}^{\alpha}u_{(j)}^{\beta}\rangle .
\label{equation16}
\end{equation}
where $V$ is the body's volume,  $E_0=\int d{\bf r}_0 \epsilon_0 $ is the rest energy of each unit cell and the angular brackets denote an average over all units cells with $\gamma_{(j)}=\gamma_{u_{(j)}}$ for short. Notice that the last term of equation(\ref{equation16}) contains the average unit cell relativistic stress and momenta over the whole body which  must vanish in the centre of inertia frame and also in accord with $p=0$ for microscopic consistency.  Thus the only non-vanishing component is
\begin{equation}
{\cal E}_0 = \int d{\bf r} T^{00}=VE_0\langle\gamma_{(j)}\rangle,
\label{equation17}
\end{equation}
a result that can also be checked from the invariant trace of the tensor $\int d{\bf r} T^{\alpha}_{\ \alpha}$ as both quantities are the only invariant quantities in this case \cite{NoteB}.  Let us show that this quantity transform as a four vector.  Under the Lorentz transformation of equation(\ref{equation2}), we have:
\begin{eqnarray}
P^0 = {1\over c}\int d{\bf r} T^{00}  &=& {\gamma^2\over c} \int d{\bf r} [T'^{00}+2\beta T'^{01}+\beta^2 T'^{11} ] \ {\rm while}\cr
P^1= {1\over c} \int d{\bf r} T^{01}  &=& {\gamma ^2 \over c}\int d{\bf r} [(1+\beta^2) T'^{01}+\beta  T'^{00} + \beta T'^{11}].
\label{equation18}
\end{eqnarray}
Here the primes denote the quantities in the centre of inertia frame and $\beta$ and $\gamma$ are also referred to that frame.  It is now straightforward following the arguments leading to equation (\ref{equation17}) to once again that the above quantities transform as a true-vector as follows:
\begin{eqnarray}
P^0 = {\gamma^2\over c} \int d{\bf r} T'^{00}&=&{\gamma\over c}{\cal E}_0 \ {\rm while}\cr
P^1 = {\gamma ^2 \over c}\int d{\bf r} \beta  T'^{00}&=&{\gamma \beta\over c}{\cal E}_0  .
\label{equation19}
\end{eqnarray}

\subsection{Stressed body}
For a body under stress we must introduce Poincar\'e stresses to provide mechanical stability.  One way to do this is to introduce an electric field as before and the energy-momentum tensor is now:
\begin{equation}
T^{\alpha\beta}= (p+\epsilon_0) u^{\alpha}u^{\beta}+ 2 W_E \tau^{\alpha\beta}-(p+W_E)g^{\alpha\beta},
\label{equation14b}
\end{equation}
where once again the field is taken along the $x=x_1$ axis and here $\tau^{\alpha\beta}$ is the global diagonal tensor:
\begin{equation}
\tau^{\alpha\beta}=
\begin{pmatrix}
1&0&0&0\cr
0&-1&0&0\cr
0&0&0&0\cr
0&0&0&0 \\
\end{pmatrix}.
\label{equation14c}
\end{equation}
Now all the quantities in equation(\ref{equation18}) for this case can be evaluated as before. However we shall postpone the details to the appendix where it will be shown that:
\begin{eqnarray}
\int d{\bf r} T'^{00}={\cal E}_{0W} &=& {\cal E}_0 + {\cal E}_W \ {\rm and} \cr
\int d{\bf r} T'^{01} &=& \int d{\bf r} T'^{11}=0,
\label{equation20}
\end{eqnarray}
verifying once again equation (\ref{equation19}) where ${\cal E}_0$ is now replaced by ${\cal E}_{0W}$ which must now include the static field: ${\cal E}_W= V E_W <\gamma_{(j)}>$, see appendix.

Finally we shall comment on the robustness of our derivation which is no surprise.  Following the procedure introduced by Rohrlich \cite{Rohrlich1,Hnizdo1,Moylan,Jackson2}, we can treat the static $E$ field or Poincar\'e stress separately, by first using the results of sub section \ref{stressless body} and then introduce the electromagnetic tensor $T_E^{\alpha\beta}$ in a manifestly covariant prescription, which guarantees a true four vector $P_E^{\alpha}$ that can be added to equation(\ref{equation19}).  This is given by \cite{Rohrlich1,Hnizdo1,Jackson2}:
\begin{equation}
P_E^{\alpha}= {\gamma\over c} \int  d{\bf r}\ u_{\beta}\ T_E^{\alpha\beta},
\label{equation21}
\end{equation}
where $u_{\beta}=\gamma(1,\beta,0,0)$ is the four velocity defining our Lorentz transformation of equation(\ref{equation2}) and $T_E^{\alpha\beta}$ is the external field in equation(\ref{equation8}) namely :
\begin{equation}
T_E^{\alpha\beta}=
\begin{pmatrix}
W_E&0&0&0\cr
0&-W_E&0&0\cr
0&0&W_E&0\cr
0&0&0&W_E \\
\end{pmatrix}.
\label{equation14d}
\end{equation}
The same result as that derived above via the Klein-Abraham definitions would follow.  This approach is more powerful and will certainly be useful for the consideration of radiation as an extension of the results of subsection \ref{CapacitorwithRadiation}.

\section{Conclusion}

In conclusion, we have provided an alternative derivation of the Klein-Abraham's theorem and Einstein's signature relation from the macroscopic perspective, including the case where the body contains electromagnetic radiation. Diagonalisation of the energy-momentum tensor is an important theme that is not restricted to macroscopic models and should be useful for future research such as in microscopic models. This work also demonstrates the minor role of Poincar\'e stresses and supplements the earlier results presented by previous authors in this journal \cite{Ohanian1,Romer1,Holstein, Rohrlich1,MacDonald,Hnizdo1,Hnizdo2,Rohrlich2,Moylan} which should be included in future revisions of standard textbooks on the subject.  However, we must also conclude by a caution against accepting this work as a definitive proof of $E_0=mc^2$ for a relativistic {\it point} particle. A key step in our proof requires the use of eqn(8) of Ohanian\cite{Ohanian1} and its comparison with the relativistic momentum of a free point particle: $m{\bf v}/\sqrt{(1-{v^2\over c^2}})$.  Since the latter is a fundamental result of relativistic Lagrangian dynamics, and in the rest case $(v=0)$ is what one would be setting out to prove in the case of a point particle, then its use in the proof becomes a tautology.  As such the relation $E_0=mc^2$ for a relativistic {\it point} particle must be considered unproven and must be empirically based.

\section{Appendix}
The integral for the $T'^{01}$  component in equation(\ref{equation18}) can be evaluated from:
\begin{equation}
\int d{\bf r} T'^{01} = (\sum_j\gamma_{(j)}u_{(j)}^1 )\int d{{\bf r}_{(j)}}[(p+\epsilon_0)]=0 ,
\label{equation22}
\end{equation}
where the zero result follows as the cell averaged relativistic momentum vanish in the centre of inertia frame as before, see also below. The second integral we need is for the $T'^{11}$  component in equation(\ref{equation18}) which reduces to:
\begin{equation}
\int d{\bf r} T'^{11} = \Bigr(\sum_j\gamma_{(j)}(u_{(j)}^1)^2 \int d{{\bf r}_{(j)}}[(p+\epsilon_0)]\Bigl)-\int d{\bf r}(W_E-p)=0,
\label{equation23}
\end{equation}
where the last integral vanishes due to our added Poincar\'e stress $p=W_E$ and the first can be taken out of the sum which vanishes as the cell averaged relativistic pressure must be zero in the centre of inertia frame in accord with $p=W_E$ for mechanical stability, see sub section \ref{section B}.  Finally we shall evaluate the total energy term i.e. the integral for the $T'^{00}$ which can easily be shown to reduce to: \begin{equation}
\int d{\bf r} T'^{00} = \Bigr(\sum_j\gamma_{(j)}\int d{{\bf r}_{(j)}}[(p+\epsilon_0)]\Bigl)+\int d{\bf r}(W_E-p)=\sum_j \gamma_{(j)}\int d{{\bf r}_{(j)}}[p+\epsilon_0],
\label{equation24}
\end{equation}
where the integral over $W_E-p$ again vanishes, so that our final result is given by:
\begin{equation}
\int d{\bf r} T'^{00} = \sum_j \gamma_{(j)}\int d{{\bf r}_{(j)}}[\epsilon_0+W_E]=(E_0+E_W)V <\gamma_{(j)}>.
\label{equation25}
\end{equation}


\begin{thebibliography}{10}
\bibitem{Landau2}  Reducing the tensor $T^{\alpha\beta}$ to diagonal or block diagonal form, and a  consideration of the various cases involving material and radiation, is an important exercise that cannot be over emphasized, see for example reference \cite{Landau1} sec 35 and 94.
\bibitem{NoteA} The literature, including this journal, is dotted with references to Poincar\'e stresses, which are precisely such stresses needed to obtain stability, see for example \cite{Hnizdo3} and an alternative manifestly covariant prescription pioneered by Rohrlich \cite{Rohrlich2} which eliminates such stresses, see later.
\bibitem{Landau4} See reference \cite{Landau1} pg 80--87.
    This reference also includes a treatment using a classical virial theorem and another giving the equation of state for a non-interacting gas plus radiation through the tensor invariant $T^{\alpha}_{\ \alpha}$ or its time average.
\bibitem{Landau5} See footnote on pg 85 in reference\cite{Landau1}. This is strictly not necessary but it simplifies out maths.
\bibitem{NoteB} Our result parallels that of reference\cite{Landau1} (cf equation (35.9) in section 35), in which they relate the unit cell quantities to the microscopic model of a non-interacting ideal gas.  However in that case, it is a hypothesis, while here it is dictated by the relativistic macroscopic description equation(\ref{equation14}).
\bibitem{Jackson2} This prescription is now formally discussed in some textbooks in relation to the stability of the classical electron, see for example \cite{Jackson} sec 16.4, but it still remains relatively unknown in terms of applications. It appears Dirac was one of the the first to notice that one can ignore the stabilising force in a manifestly covariant treatment, see appenidx 4 in \cite{Rohrlich0}.
\end{thebibliography}
\end{document}